# An empirical review of the different variants of the Probabilistic Affinity Index as applied to scientific collaboration

Zaida Chinchilla-Rodríguez

*Instituto de Políticas y Bienes Públicos (IPP), Consejo Superior de Investigaciones Científicas (CSIC), Madrid 28037, Spain*

Yi Bu\*

*Department of Information Management, Peking University, Beijing 100871, China*

Nicolás Robinson-García

*Delft Institute of Applied Mathematics, TU Delft, 2628 XE Delft, Netherlands.*

Cassidy R. Sugimoto

*Luddy School of Informatics, Computing, and Engineering, Indiana University, Bloomington, IN 47408, U.S.A.*

\* Correspondence concerning this article should be addressed to Dr. Yi Bu (buyi@pku.edu.cn).

**Abstract:** Responsible indicators are crucial for research assessment and monitoring. Transparency and accuracy of indicators are required to make research assessment fair and ensure reproducibility. However, sometimes it is difficult to conduct or replicate studies based on indicators due to the lack of transparency in conceptualization and operationalization. In this paper, we review the different variants of the Probabilistic Affinity Index (PAI), considering both the conceptual and empirical underpinnings. We begin with a review of the historical development of the indicator and the different alternatives proposed. To demonstrate the utility of the indicator, we demonstrate the application of PAI to identifying preferred partners in scientific collaboration. A streamlined procedure is provided, to demonstrate the variations and appropriate calculations. We then compare the results of implementation for five specific countries involved in international scientific collaboration. Despite the different proposals on its calculation, we do not observe large differences between the PAI variants, particularly with respect to country size. As with any indicator, the selection of a particular variant is dependent on the research question. To facilitate appropriate use, we provide recommendations for the use of the indicator given specific contexts.

## 1. INTRODUCTION

Scientific collaboration is the dominant mode of production across most scientific disciplines (Adams, 2013; Sugimoto & Larivière, 2018; Chinchilla et al., 2018). The rise in collaboration across the last century (Lariviere et al., 2015) has been accompanied by a concomitant interest in the study of collaboration practices (e.g., Beaver & Rosen, 1978, 1979; Leahey, 2017; Price, 1986; Sonnewald, 2007). These studies have focused on three main factors for and consequences of collaboration, namely economic, e.g., the need of sharing infrastructures and costs (Beaver & Rosen, 1979; Price, 1986); cognitive (Beaver & Rosen, 1978; Edge, 1979; Stokes & Hartley, 1989); and social factors, e.g., geopolitical, historical, linguistic, and cultural similarity (Luukkonen et al., 1992).

In this study, we explore the Probabilistic Affinity Index (PAI), an indicator used to identify the preferred partners of a given country in country-country co-authorship networks (e.g., Chinchilla et al., 2018). According to Zitt et al. (2000), "t(T)his indicator was designed to account for several factors that drive co-authorship—such as cultural and geographical proximities in which common historical experience plays a central role" (p. 629)… "Considering all potential partners of a country, PAI reveals the association strength of countries in terms of propensities, intensities, or affinities in collaboration linkages, which are largely dependent upon different drivers of scientific collaboration. The indicator reveals the limitations of size-dependent indicators and is particularly useful for highlighting small relationships that might be obscured by other indicators" (p. 633).

Differences in the volume of single and multi-authored papers create some issues when people study scientific collaboration (Luukonen et al., 1993). To go beyond absolute differences in country sizes and estimate 'propensities' or 'intensities' of collaboration, it is necessary to develop measures which take size into account. Different solutions have been proposed in the form of similarity measures for normalization purposes within the field of scientometrics. Despite continued debate, there is some consensus on the use of probabilistic similarity measures such as PAI, —in which association strengths are measured,—instead of theoretic similarity measures (e.g., cosine, inclusion index, or Jaccard index) (van Eck & Waltman 2009). However, there is neither a single formula nor a single index for applying this algorithm and it is difficult to replicate or discriminate among variants. In extant literature, three major algorithms have been proposed to define PAI, namely non-overlapping, overlapping, and self-exclusive definitions, respectively. The purpose of this paper is to review each of them, to show how they are computed, to examine each variant, and to explain their differences based on the analysis of a set of countries. For this, we structure the paper in two main parts; a conceptual one in which we review previous literature, and an empirical one in which we compute and compare the different variants for a selection of countries. We then discuss similarities and differences and conclude with a series of recommendations on the selection of a PAI variant based on the purpose of the study.

## 2. LITERATURE REVIEW

In this section we review studies on the use of indicators for assessing international collaboration, focusing on the different PAI variants. First, we provide a basic overview on the main considerations when analyzing scientific collaboration, such as the use of co-authorship and counting methods employed. Then, we revise the historical roots of the PAI. We conclude this section by discussing the main differences between PAI variants.

*2.1. Main considerations for analyzing scientific collaboration in scientometrics*

Scientometric studies typically use co-authorship as a proxy of international collaboration between scientometric entities, e.g., individuals, institutions, and countries. However, it should be noted that not all collaborative efforts end up in publications, co-publishing is not the sole reason for collaborating (Katz & Martin, 1995), and scientometric studies cannot fully explain the dynamics of scientific collaboration (Heinze & Kuhlmann, 2008). Despite this, co-authorship constitutes one of the best-documented and studied evidences of relationships among researchers, institutions, and countries funding and/or conducting research (Bordons & Gómez, 2000; Glänzel & Schubert, 2004; Velden, Haque, & Lagoze, 2010).

Another important issue when analyzing international collaboration is the counting method used. The problem of how to distribute credit for authorship has been extensively studied since the early developments of authorship indicators (e.g., Frandsen & Nicolaisen, 2010; Harsanyi, 1993; Lindsey, 1980; Waltman, 2016). Recently, Gauffriau (2017) published a review of counting methods and arguments for use. Among these, the most popular approaches are full and fractional counting methods. In cases in which researchers use small datasets, full and fractional counting do not make fundamental differences when a co-authorship network is constructed. Nevertheless, using distinct strategies with larger datasets might yield in quite different conclusions and implications (e.g., Perianes-Rodriguez et al., 2016).

The advantages and disadvantages of the two methods have been well-discussed (e.g., Braun, Glänzel, & Schubert, 1991; Okubo, Miquel, Frigoletto, & Doré, 1992). For instance, compared with full counting, a fractional counting strategy could prevent misunderstandings or misinterpretation, particularly when comparing output across fields. However, an increase of internationally co-authored publications could lead to an overall decline on the number of papers (Leydesdorff, 2001). This makes it hard to interpret productivity when using fractional counting, as "non-integer weights" are employed (Perianes-Rodriguez et al., 2016), leading to counterintuitive results in some cases (Park et al., 2016). Also, as an analysis at the international level would lead to a zero-sum game, internationalization would be seen as a negative factor when analyzing the performance of collaboration programs (Leydesdorff, 1988).

A stream of literature argues in favor of the full counting method in international collaboration. According to Okubo et al. (1992), full counting helps to observe the volume of contacts created by scientists, considering a contact as a link that always has the same value between any two (or more) countries, regardless of the number of participants. Leclerc and Gagne (1994) prefer full to fractional counting:

"What we feel must truly be measured is less the overall production of collective scientific work than the actual contribution or participation of national scientific communities in the knowledge construction cycle. But international scientific cooperation, the coauthored articles of which reflect the scope as well as the real extent of exchange networks, has a decisive function and growing share of scientific activity" (p. 264)

Thus, it can be argued that international collaboration should be considered as an achievement on all involved actors, and should thus be honored with a full counting strategy (Park et al., 2016).

Both methods involve biases that should be kept in mind when using one or the other. The final decision of selecting which counting strategy will be determined by the research question addressed. Most of the previous studies employing PAI utilized a full counting strategy except Leclerc and Gagne (1992) and Zitt et al. (2000), who argued that the two methods yielded similar results.

*2.2. History of the Probabilistic Affinity Index (PAI) and Related Measurement Studies*

Within the field of scientometrics, PAI has been labeled under different denominations and applied with slight differences. The key point in most studies is whether an indicator reflecting affinity between countries should be sensitive or not to the size of countries in the network, as well as geographical and historical factors. In Table 1 we briefly summarize the main studies which we consider to be milestones on the development of PAI and its different variants in scientometrics.

**Table 1. Main milestones and proposition of affinity indices**

| Reference | Purpose of the study | Contribution |
| --- | --- | --- |
| Frame & Carpenter (1979) | Analyze international collaboration worldwide | Note the relation between national size and proportion of internationally co-authored papers |
| Schubert & Braun (1990) | Study the intrinsic cooperativity of countries | Propose a "cooperation index" which corrects for country size |
| Luukonen et al. (1992) | Propose a method to measure affinity between countries regardless of their size | Introduce the "exclusive strategy" for asymmetrical values |
| Leclerc & Gagné (1994) | Describe the components of world science by using bibliometric indicators of collaboration | Introduce the Weighted Affinity Index |
| Zitt et al. (2000) | Characterize the collaboration profiles of France, Germany, United Kingdom, United States and Japan | Define the PAI as a ratio of observed to expected values of collaboration normalized by the size of both countries becoming a non-size dependent indicator. |
| Yamashita & Okubo (2006) | Propose the Probabilistic Partnership Index (PPI) and compare it with the PAI and other indicators | Introduce PPI as a standardized ratio between the observed and expected number of links, providing another view of deviation of the expected value |

The first study on collaboration using co-authorship to highlight the influence of national scientific systems' size and the proportion of internationally co-authored publications was one by Frame and Carpenter (1979). In their study on international collaboration, they proposed an index which was then used by Schubert and Braun (1990) to establish a general relationship between the number of total and foreign co-authored publications. This index was used to characterize individual countries by their deviations from the general trend.

Luukkonen et al. (1992) examined which factors influence co-authorship linkages among the 30 most prolific countries in terms of number of scientific articles for the 1981-1986 period. They used two formulas to calculate the observed/expected ratio of co-authorship for each pair of countries, indicating their collaboration preferences (pp. 107, 114). For symmetrical values: $(C_{x,y} * \mathrm{T})/(C_x * C_y)$, where $C_{x,y}$ refers to the number of collaborations between countries $x$ and $y$, $C_x$ total number of collaborations country $x$ has with other countries considered, $C_y$ total number of collaborations country $y$ has with other countries considered, and T total number of collaborations among all countries considered in the network. For asymmetrical value [1]: $[C_{x,y} * (T - C_x)]/(C_x * C_y)$ .[2] The asymmetrical value is referred to as "exclusive strategy". This strategy was further supplemented by Schubert and Glänzel (2006), who referred to it as "co-authorship preference index."

Luukkonen et al. (1992) conducted a Correspondence Factorial Analysis and a Minimum Spanning Tree analysis to study countries' involvement in international scientific networks. In addition, Luukkonen et al. (1993) provided three more possible ways to normalize the relative strength: bilateral similarity measures (e.g., Salton's and Jaccard measurements), multilateral similarity measures (e.g., Goodman's quasi-independence model), and multidimensional scaling methods. Most of these strategies are found to help researchers better interpret the results obtained from relative strength measurements. However, some of these measurements tend to underestimate the strength of links between small countries with a low scientific output (Luukkonen et al., 1992, p. 24). The goal of PAI was to account for this. PAI is interpreted as follows: If the value of PAI of two countries is above one, then there are more scientific collaborations between those two countries than expected, given their size and tendency to collaborate internationally. The formulas overemphasize countries that have a highly skewed distribution of collaborations with one or two dominant partners. To illustrate this, Luukonen et al (1992, p. 115) provide the following example:

> "Many scientists from country A go to country B for doctoral studies. This eventually results in many papers coauthored with supervisors in country B. If scientists in country A have very few scientific contacts with countries other than B, B features as the most important collaborative partner for A. As far as B is concerned, A seems important, since A's collaborations are highly concentrated with B and, therefore, the observed values far exceed those expected. In addition, the fact that those co-authorship matrices are sensitive to variations that depend on which countries are included or excluded from the analysis is also highlighted. If an important partner or partners of a country have been excluded from the analysis, then this country's position will change."

---

[1] Leclerc and Gagne (1994) named this index as Weighted Affinity Index (WAI) that indicated the relative and mutual affinity between countries (p. 267).
[2] The annotations here refer to the same items M7 indicates in the later sections.

Leclerc and Gagné (1994) employed the proximity index (PRI) to measure the intensity of scientific exchanges between two countries based on the number of co-authored publications measured and collaborations theoretically expected. Under this idea, they defined the proximity index between countries $i$ and $j$: $PRI(i,j) = \frac{C_{i,j} \cdot T}{C_i \cdot C_j}$, where $C_{i,j}$ = number of articles co-authored internationally by countries $i$ and $j$, $T$ = total number of articles co-authored internationally by all countries in a given dataset, $C_i$ ($C_j$) = total number of articles co-authored internationally by country $i$ ($j$) with all other countries in the database. PRI with values larger than one reflects "higher collaboration intensity between two countries than their respective weight and propensity to collaborate would indicate" and, therefore, "shows the symmetry of relations between the countries." (p. 266). Since all definitions in Leclerc and Gagné (1994) are based on paper count, there should not be any overlapping of co-author relationships; we therefore name their algorithm as a "non-overlapping" method.

Zitt et al. (2000), on the other hand, used the probabilistic affinity index (PAI) in a similar way, but their $C_i$ ($C_j$) is defined as the total number of "co-authorship linkages" of country $i$ ($j$). The major difference between these two approaches is that while the latter used co-authorship-level counting, the former uses publication-level counting. The number of co-authorships in internationally collaborative publications is often much more than that of international co-authorship links (Okubo et al., 1992). Compared with the non-overlapping method, Zitt et al. (2000) showed how to calculate

PAI under an "overlapping" circumstance. In addition, this paper introduces a key aspect in its calculation. The index may be defined with or without auto-co-authorships (here, auto-authorships means collaborations within the country, normally the values in the diagonals of the international collaboration matrix). Keeping the diagonal makes the index dependent on the propensity of countries to internal co-authoring, which varies greatly among countries, and may bias comparisons. They used an iterative process of recalculation of margins for neutralizing the auto-co-authorship, so that the final value in the diagonal becomes neutral (order-zero reconstitution) (p. 633). This strategy has also been mentioned in National Science Foundation[3] (2012) though in a mathematically different style.

Yamashita and Okubo (2006) measured inter-sectoral cooperation between France and Japan. They combined PAI with Salton's index as a measurement to indicate the relative strength between two countries $i$ and $j$. That is, $r_{ij} = \frac{n_{ij}}{\sqrt{n_i n_j}}$, where $n_{ij}$ refers to the number of co-authored papers between countries $i$ and $j$ and $n_i$ ($n_j$) the total number of papers published by the country $i$ ($j$). Note that this measurement is a variant of the so-called *Ochiai* coefficient (Ochiai, 1957; Zhou & Leydesdorff, 2016). Nonetheless, there are still debates on how these normalization strategies should be proposed in different circumstances (e.g., Ahlgren, Jarneving, & Rousseau, 2003, 2004; van Eck & Waltman, 2009; White, 2003).

Moreover, Yamashita and Okubo (2006) introduced the probabilistic partnership index (PPI), a standardized ratio between the observed and expected numbers of links, providing another view of deviation from the expected value. PPI is formulated as follows: $PPI = (nij) - \frac{Ea[nij]}{\sigma}$, where $Ea[nij]$ and $\sigma$ are the expected value and standard deviation in the distribution of the number of links between sectors $i$ and $j$ under the constraint of the number of articles and current participants, which is estimated by the Monte-Carlo method. PPI is a standard score of $nij$ against the probability distribution of current participants without any preference for collaborating partners (p. 308). Yamashita and Okubo (2006) also compared PPI with Jaccard, Salton-Ochiai, and PAI indexes, and pointed out that PAI measures the relative strength of each co-operative link in comparison with the total linkage, while PPI measures the rareness of occurrence of the observed value, in comparison with an assumed distribution of links generated by randomly distributing participants within the current articles.

---

[3] https://www.nsf.gov/statistics/seind12/c0/c0s7.htm

*2.3. Main differences between PAI variants*

Based on the historical account on the different propositions of indicator, we observe that all variants revolve around two specific steps on the calculation of the PAI, that is, 1) processing of the main diagonal of the co-authorship matrix, and 2) normalization strategy.

With regard to the processing of the co-authorship matrix, all variants consider that all PAI values among countries listed in the co-authorship matrix should be contained in a new matrix. But they differ on the main diagonal of this new matrix, $PAI(i,i)$. As mentioned above, in previous PAI studies, we have found at least two ways to address this problem: (1) set the main diagonal values to empty or zero values (e.g., Luukkonen et al., 1992; Leclerc & Gagne, 1994; Schubert & Glänzel, 2006; Chinchilla-Rodríguez et al., 2017); or (2) use an iterative strategy to automatically fill in the main diagonal values (Zitt et al., 2000). Finardi and Buratti (2016, p. 438) detailed how an iterative strategy should be used in PAI main diagonal processing. They suggest that the final output can be defined as the total number of papers published by each country, in international collaboration, or only domestic publications[4].

---

[4] Inspired by Eom (2008), six approaches could been used in processing main diagonal values in raw co-citation

In the case of normalization strategies followed, Zitt et al. (2000) applied an easy way to implement the normalization of PAI. The normalized PAI between countries $i$ and $j$, $N\_PAI(i,j)$, is calculated as $N\_PAI(i,j) = \frac{PAI(i,j)^2 - 1}{PAI(i,j)^2 + 1}$, although a later study applied it without power $N_{PAI(i,j)} = \frac{PAI(i,j) - 1}{PAI(i,j) + 1}$ (Yamashita & Okubo, 2006). By doing either of these, PAI is normalized between -1 and 1, and a zero value of $N\_PAI(i,j)$ indicates a neutral statement (i.e., a baseline/standard). Some studies have followed this normalization strategy (e.g., Chinchilla-Rodríguez et al., 2017; Yamashita & Okubo, 2006), although there are other cases in which no normalization procedure was applied (e.g., Schubert & Glänzel, 2006).

## 3. DATA AND METHODS

In the current study, we aim to show how using different variants of PAI would affect the results. We first set up a framework on how to use PAI in international scientific collaboration studies with a four-step procedure, namely 1) data acquisition and preprocessing, 2) raw PAI matrix construction, 3) normalized PAI matrix transformation, and 4) results visualization and interpretations.

---

matrices in author co-citation analysis (ACA) that aims to depict scientific intellectual structures: (1) Missing values (McCain, 1990; White & McCain, 1998); (2) The mean co-citation count for each author (McIntire, 2007); (3) Zero (e.g., Bu, Liu, & Huang, 2016; Bu, Ni, & Huang, 2017); (4) Highest off-diagonal cocitation counts; (5) Three highest off-diagonal values divided by two (White & Griffith, 1981); and (6) Raw cocitation frequency. Similar to ACA, PAI studies also needs a diagonal matrix containing co-authorship frequencies. Therefore, we believe that these strategies should be considered to use and fit in process PAI main diagonal issues in the future studies; but determining to use a specific strategy, once again, should depend on what research questions are going to be addressed.

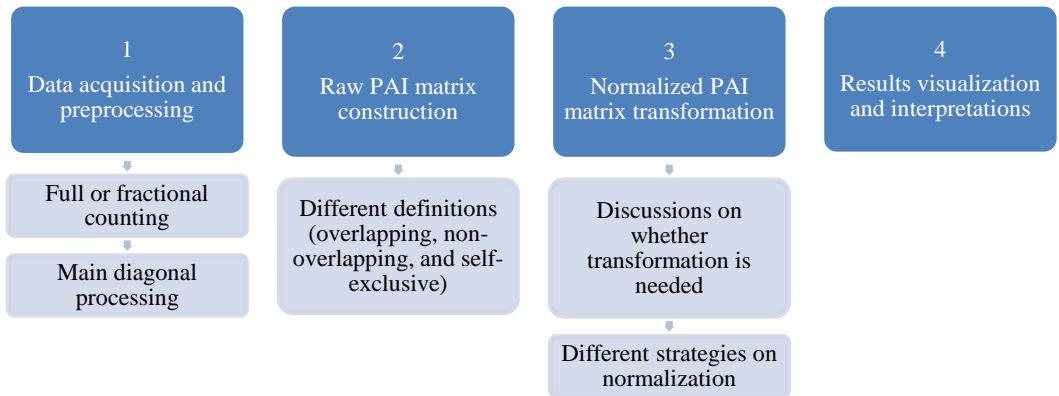

**Figure 1. A four-step framework of using PAI in international scientific collaboration studies.**

*3.1. Data acquisition and preprocessing*

The first step needed is to collect the data and preprocess it prior to analyzing it. The expected output of this step is to construct a matrix containing the number of international scientific collaborations among all country/region pairs, that is, the co-authorship matrix. We annotate the number of countries/regions in the dataset as $N$, [5] and the number of internationally collaborative publications between countries $i$ and $j$ as $n_{ij}$. Before constructing the co-authorship matrix, one should determine the counting strategy to follow. If full counting is used, this matrix will simply contain all values of $n_{ij}$. Another thing worth noticing is the main diagonal values of this matrix, as discussed in Section 2.

The dataset used in the current study was derived from the CWTS (Leiden University) in-house version of the Web of Science (WoS) database. 13,699,176 distinct records between 2008 and 2015 were selected. Authors' names were disambiguated based on Caron and van Eck (2014), after which 15,931,847 unique authors were identified, where ~3.7% are affiliated to more than one country. Country names were cleaned and disambiguated, resulting in a total of 213 countries. We then constructed a 213*213 international scientific collaboration matrix; each of the elements in this matrix equals the number of papers co-authored between each pair of countries.

---

[5] Although most "countries" contained are really countries, some are regions/areas, e.g., Taiwan. In this paper, however, for the sake of simplicity, we just name them as "countries" in the following sections. But note that they actually mean "countries/regions."

In this paper, we applied a full counting strategy for simplicity and better interpretation. In terms of the main diagonal strategy, this paper provides two approaches. The first is just setting all main diagonal values to zero (e.g., Leclerc & Gagné, 1994; Chinchilla-Rodríguez et al., 2017). The second approach employs an iterative strategy, as mentioned in Finardi and Buratti (2016): specifically, we set all main diagonal values to zero at the beginning; then each value of the main diagonal, $n(i,i)$, is calculated as $n(i,i) = \frac{\sum_j n(i,j)}{\sum_k \sum_j n(j,k)}$; once values of $n(i,i)$ no longer change in a new round of calculation, we stop the iteration. The collaboration matrix was iterated 29 times.

*3.2. Raw PAI matrix construction*

Next, we constructed a raw PAI matrix construction following each of the methods. At this stage, two main issues should be discussed. First, we should consider which PAI is defined. We here apply three different branches of PAI definitions in the current study, namely non-overlapping, overlapping, and self-exclusive strategies, respectively. For each type of PAIs, there may be more than one similar definition but with slight differences.

Given a dataset containing $N$ countries, suppose that the total number of internationally collaborative publications is $n_{all}$, the total number of values in the co-authorship matrix is $n(...)$, the number of internationally collaborative papers between countries $i$ and $j$ is $n_{ij}$, and $n_i$ ($n_j$) represents the total number of internationally collaborative papers of country $i$ ($j$). We here define all PAIs based on their categories, as shown in Table 2.

**Table 2. Annotations and types for different PAI definitions.**

| Method | Annotation | Type | Reference |
|---|---|---|---|
| M1 | Non_overlapping_0 | Non-overlapping | Leclerc & Gagné (1994) |
| M2 | Overlapping_0 | Overlapping | Zitt et al. (2000) |
| M3 | Overlapping_iterative | Overlapping | Proposed by Zitt et al. (2000) and detailed in Finardi & Buratti (2016) |
| M4 | Overlapping_all | Overlapping | Zitt et al. (2000) |
| M5 | Overlapping_inter | Overlapping | Zitt et al. (2000) |
| M6 | Overlapping_intra | Overlapping | Zitt et al. (2000) |
| M7 | 1992_variant_P114 | Self-exclusive | First proposed by Luukkonen et al (1992) and applied by Schubert & Glänzel (2006) |

The non-overlapping PAI refers to Leclerc and Gagné (1994) (annotated as M1). Specifically, it is calculated as:

$$M1(i,j) = \frac{n_{all} * n_{ij}}{n_i * n_j}$$

Note that M1 defines the main diagonal of its PAI matrix as zero. Therefore, their annotation in Table 2 ends with a "_0" (the same below).

The second branch, namely overlapping PAIs, is defined by Zitt et al. (2000) but there are five different variants due to distinct main diagonal definitions (annotated as M2-M6):

$$Mr(i,j) = \frac{n(...) * n_{ij}}{\sum_k n(i,k) * \sum_k n(j,k)}$$

where $r = 2,3,4,5,6$. In M2, all main diagonal values are set as zero in advance. In M3, we employ the iterative strategy proposed by Finardi and Buratti (2016) to determine the values. In M4-M6, the main diagonal values are set as the country's total publication count, total international collaboration count, and total intra-country collaboration count, respectively. All these strategies (M2-M6) have been mentioned and discussed in Zitt et al. (2000). NSF (2012) proposed a seemingly different strategy but mathematically it is the same as M2-M6.

Here we refer to the self-exclusive PAI as M7. This method was first proposed in the Page 114 in Luukkonen et al. (1992) but applied by Schubert and Glänzel (2006)[6]:

$$M7(i,j) = \frac{n_{ij} * (n(...) - \sum_k n_{ik})}{\sum_k n_{ik} * \sum_k n_{jk}}$$

Essentially M7 is also a variant of the overlapping strategy but possesses a self-exclusive feature because we can find that in the nominator of M7, $\sum_k n_{ik}$ is subtracted.

*3.3. Normalized PAI matrix transformation*

For the non-overlapping and overlapping approaches, we use Zitt et al. (2000)'s method to normalize the raw PAI result. The normalized PAI for all of the three approaches is calculated as follows:

$$NPAI(i,j) = \frac{PAI(i,j)^2 - 1}{PAI(i,j)^2 + 1}$$

In the case of the M7 approach (self-exclusive), we did not apply any normalization strategy but followed the instructions detailed in Schubert and Glanzel (2006).

*3.4. Results visualization and interpretation*

The final step after calculating and normalizing PAIs is to visualize and interpret results. While it is possible to rank the PAI of any country, it is not advisable due to the bias small countries might introduce. Alternatively, we followed the strategy provided in previous studies (Chinchilla-Rodríguez et al., 2018), in which we select a target country and then calculate the Affinity index (AFI) for each country pair, rank all countries in a descending order, select the top n countries, and rank these *n* countries by their PAI values. This strategy helps to identify countries where it makes sense to analyze their international scientific collaborations using the PAI values.

---

[6] Schubert and Glänzel (2006) used a seemingly-different definition of Luukkonen et al. (1992, p.114), but mathematically they are the same.

In this paper, we show as proof-of-concept the results of five countries: The United States, Spain, Singapore, Slovenia, and Kenya. These countries are located in different continents and have relatively different scientific sizes, historical roots, levels of development, expenditure in research and development (R&D) and income level (see in Table 3). We compare the PAI variants of each country by using Pearson and Spearman correlations and an R-square regression analysis.

Table 3. Basic indicators per country.

| Country | # of papers | # of internationally collaborative papers | % of internationally collaborative papers | # of collaborative countries | % of R&D investment among GDP | Population | # of papers per 1000 hab |
|---|---|---|---|---|---|---|---|
| USA | 4201368 | 1,150,830 | 27.39 | 211 | 2.72 | 318,907,401 | 13.17 |
| SPAIN | 503427 | 211,623 | 42.04 | 196 | 1.25 | 46,480,882 | 10.83 |
| SINGAPORE | 97045 | 53,726 | 55.36 | 170 | 2.14 | 5,469,724 | 17.74 |
| SLOVENIA | 33356 | 15,109 | 45.30 | 139 | 2.03 | 2,061,980 | 16.18 |
| KENYA | 12436 | 10,300 | 82.82 | 179 | 0.57 | 44,863,583 | 0.28 |

# 4. RESULTS

*4.1. Statistical comparison of PAI variants*

Table 4 shows two correlation matrices (Pearson and Spearman) for each pair of PAI variants. Overall, we observe a strong relationship among all variants of the indicator. In some cases, differences are negligible, e.g., Pearson correlations for M2-M3 and M2-M7 or Spearman for M2-M3 and M2-M5. The biggest variations in Spearman correlations are for the US, especially when M1 is compared to the rest of variants. For Pearson correlations, however, the biggest differences are found for M1-M7 in Slovenia and Kenya; and in Spain and Kenya for M3-M7, M4-M7, M5-M7 and M6-M7.

|  | Pearson | | | | | Spearman | | | | |
|---|---|---|---|---|---|---|---|---|---|---|
|  | USA | ESP | SGP | SLV | KEN | USA | ESP | SGP | SLV | KEN |
| **m1-m2** | 0.85 | 0.93 | 0.96 | 0.97 | 0.98 | 0.84 | 0.95 | 0.96 | 0.99 | 0.99 |
| **m1-m3** | 0.85 | 0.93 | 0.96 | 0.97 | 0.98 | 0.84 | 0.95 | 0.96 | 1.00 | 0.99 |
| **m1-m4** | 0.86 | 0.91 | 0.97 | 0.94 | 0.99 | 0.87 | 0.97 | 0.97 | 1.00 | 0.99 |
| **m1-m5** | 0.90 | 0.93 | 0.97 | 0.95 | 0.98 | 0.89 | 0.97 | 0.98 | 1.00 | 0.99 |
| **m1-m6** | 0.85 | 0.92 | 0.96 | 0.96 | 0.98 | 0.82 | 0.95 | 0.95 | 0.99 | 0.99 |
| **m1-m7** | 0.74 | 0.72 | 0.84 | 0.69 | 0.59 | 0.84 | 0.95 | 0.96 | 0.99 | 0.99 |
| **m2-m3** | 1.00 | 1.00 | 1.00 | 1.00 | 1.00 | 1.00 | 1.00 | 1.00 | 1.00 | 1.00 |
| **m2-m4** | 0.87 | 0.95 | 0.94 | 0.98 | 0.99 | 0.94 | 0.97 | 0.99 | 0.99 | 1.00 |
| **m2-m5** | 0.97 | 0.98 | 0.98 | 0.99 | 0.99 | 0.99 | 1.00 | 1.00 | 1.00 | 1.00 |
| **m2-m6** | 0.89 | 0.97 | 0.96 | 0.98 | 0.99 | 0.93 | 0.97 | 0.98 | 0.99 | 1.00 |
| **m2-m7** | 0.83 | 0.65 | 0.84 | 0.79 | 0.60 | 1.00 | 1.00 | 1.00 | 1.00 | 1.00 |
| **m3-m4** | 0.89 | 0.96 | 0.94 | 0.98 | 0.99 | 0.94 | 0.97 | 0.99 | 0.99 | 1.00 |
| **m3-m5** | 0.98 | 0.99 | 0.97 | 0.99 | 0.99 | 0.99 | 1.00 | 1.00 | 1.00 | 1.00 |
| **m3-m6** | 0.90 | 0.97 | 0.97 | 0.99 | 0.99 | 0.93 | 0.97 | 0.98 | 0.99 | 1.00 |
| **m3-m7** | 0.85 | 0.65 | 0.84 | 0.79 | 0.59 | 1.00 | 1.00 | 1.00 | 1.00 | 1.00 |
| **m4-m5** | 0.94 | 0.98 | 0.97 | 0.99 | 0.99 | 0.96 | 0.98 | 0.99 | 1.00 | 1.00 |
| **m4-m6** | 0.99 | 0.99 | 0.98 | 0.99 | 0.99 | 0.99 | 1.00 | 1.00 | 1.00 | 1.00 |
| **m4-m7** | 0.86 | 0.71 | 0.87 | 0.82 | 0.62 | 0.94 | 0.97 | 0.99 | 0.99 | 1.00 |
| **m5-m6** | 0.93 | 0.97 | 0.95 | 0.99 | 0.97 | 0.93 | 0.97 | 0.98 | 0.99 | 1.00 |
| **m5-m7** | 0.89 | 0.70 | 0.90 | 0.82 | 0.65 | 0.99 | 1.00 | 1.00 | 1.00 | 1.00 |
| **m6-m7** | 0.84 | 0.66 | 0.82 | 0.79 | 0.56 | 0.93 | 0.97 | 0.98 | 0.99 | 1.00 |

**Table 4. Pearson and Spearman correlations among methods. Legend: M1: non-overlapping_0; M2: overlapping_0; M3: overlapping_iterative; M4: overlapping_all; M5: overlapping_international; M6: overlapping_intra; and M7: self-exclusive.**

Figure 2 shows values (red line) and rank (blue line) for each of the five countries and combinations of methods. Overall, M2 and M7 variants are slightly higher than M1 in the U.S. However, the largest differences are observed between M3 and M4, and between M4 and M7. This means that there is a difference between choosing all publications in the matrix (M4), running the iterative process which considers the co-authorships between countries (M3), or applying the self-exclusive strategy (M7). This difference shows how the index is affected by the share of domestic publications and the size of the country in the collaboration network. This affects especially countries/regions like the United States, but also countries/regions where the percentage of domestic collaborations is higher than the international one, such as mainland China, Brazil, or Japan (Chinchilla et al., 2019). On the opposite side, we observe that when international collaboration is higher than domestic collaboration, the effect is mitigated, e.g., Slovenia or Kenya.

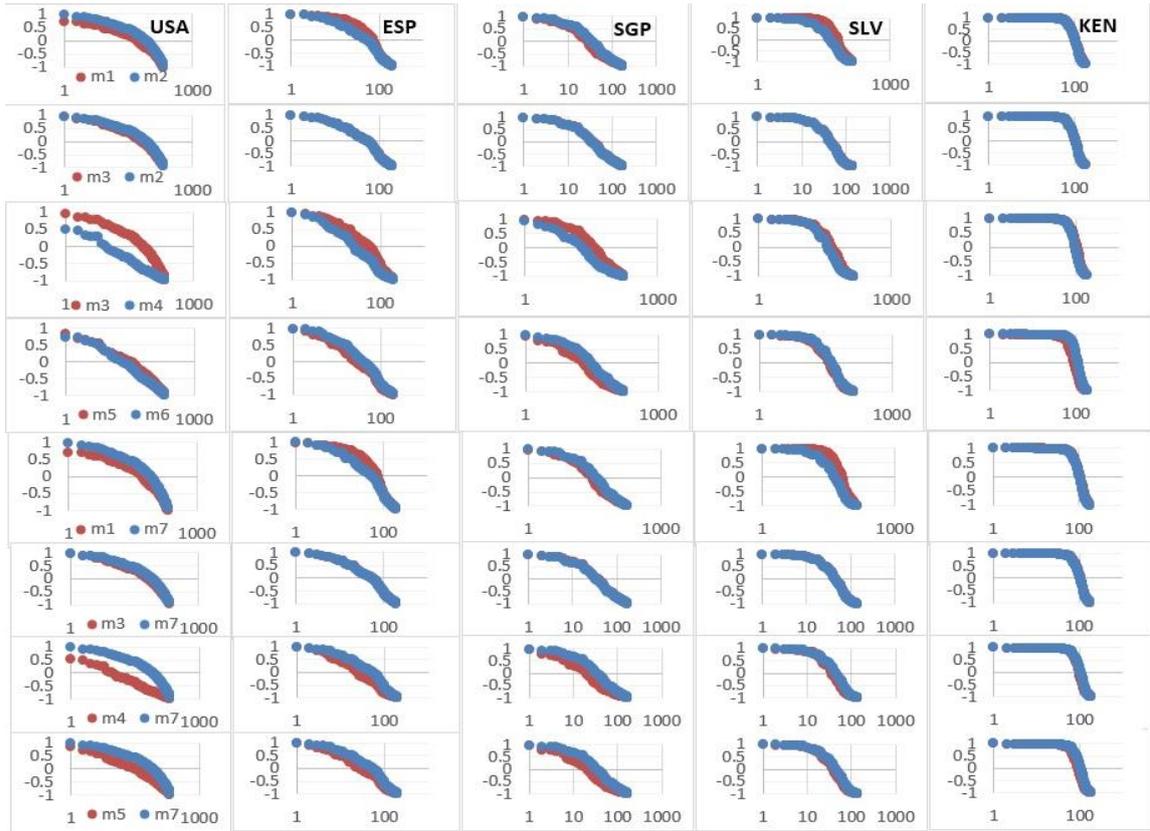

**Figure 2. Pairwise comparison of PAI variants. Overlapping of value vs. rank distributions. Only comparison of pairs with the highest differences are shown, the rest of the comparisons are available in the Supplementary Material.**

In Figure 3 we plot each pair of indicators to understand how each country is affected and by which approach. The strongest relationships ($R^2$) appear in the less prolific countries in our sample, while the lowest are for the U.S. For example, in the comparison of the non-overlapping method which considers the number of co-authored papers (M1) and the overlapping with diagonal equal to 0 (M2) considering the number of co-authorship links, there is a set of countries with higher values in the first (M1) than in the second method (M2). In the case of the United States, these countries are, for example, Republic of Georgia, Aruba, or St. Lucia (among others) with few numbers of publications and more than half of them are with international partners. In Spain, there are three countries (Equatorial Guinea, Azerbaijan and Armenia) with a high value in regard to the number of co-authored publications in contrast with the number of co-authorships. In Singapore, countries like Belize or St. Lucia prove to be preferred partners in M1 but not in M2. Overall, all these collaborators countries have a higher percentage of international collaborative papers which ranges between 65% and more than 90%).

In the comparison of overlapping with diagonal equal to zero (M2) and iterative process (M3), the results show that there is no difference between applying one or another indicator. The comparison with M3 and M4 (overlapping iterative vs. overlapping with all papers published by the country in the diagonal) demonstrates that mainland China and South Korea are preferred partners for the USA taking into account M3 but not M4. The same scenario occurs in Singapore where mainland China, Taiwan (China), South Korea, and India are the strongest partners in M3 than in M4, as well as in Slovenia with Turkey and Israel or in Kenya with the USA. That confirms that iterative process mitigates in some extent the effect of a high proportion of domestic collaboration in favor of international co-authorship links, which is corroborated with the comparison between M5 and M6.

Indeed, the use of different types of production may affect the results. For example, by using all papers to construct the matrix (M4) instead of just international papers (M5) or domestic papers (M6), we can observe how some collaborator countries/regions for the United States are considered strongest partners in M5 than in M4 (e.g., South Korea, Canada, and Israel). In the case of Spain, M5 gives more value to Brazil, Argentina, or Mexico. The same case is observed in Singapore for mainland China, Taiwan (China), or India with higher values in M5 than M4 and M6, especially for the latter two and U.S. In Slovenia, Turkey, Russia, and Italy are better positioned in M5 and, in Kenya, the U.S. and India seem more influencer partners applying M5 instead of M4. Taking into account domestic production in the diagonal (M6) instead of international collaboration (M5), the results are different especially for the smaller countries such as Slovenia and Kenya.

Finally, the comparison between M1 (non-overlapping) and M7 (self-exclusive) shows that Georgia, Armenia, and St. Lucia are stronger collaborative partners for the U.S. in M7 than in M1, contrarily to Sweden, which shows Sweden is a strong partner for the United States in M1 instead of M7. In the case of Spain, Armenia and Azerbaijan are preferred partners in M1 but not in M7. That also occurs with St. Lucia in collaborative papers with Singapore; Cyprus and Ukraine with Slovenia; Honduras and Colombia with Kenya.

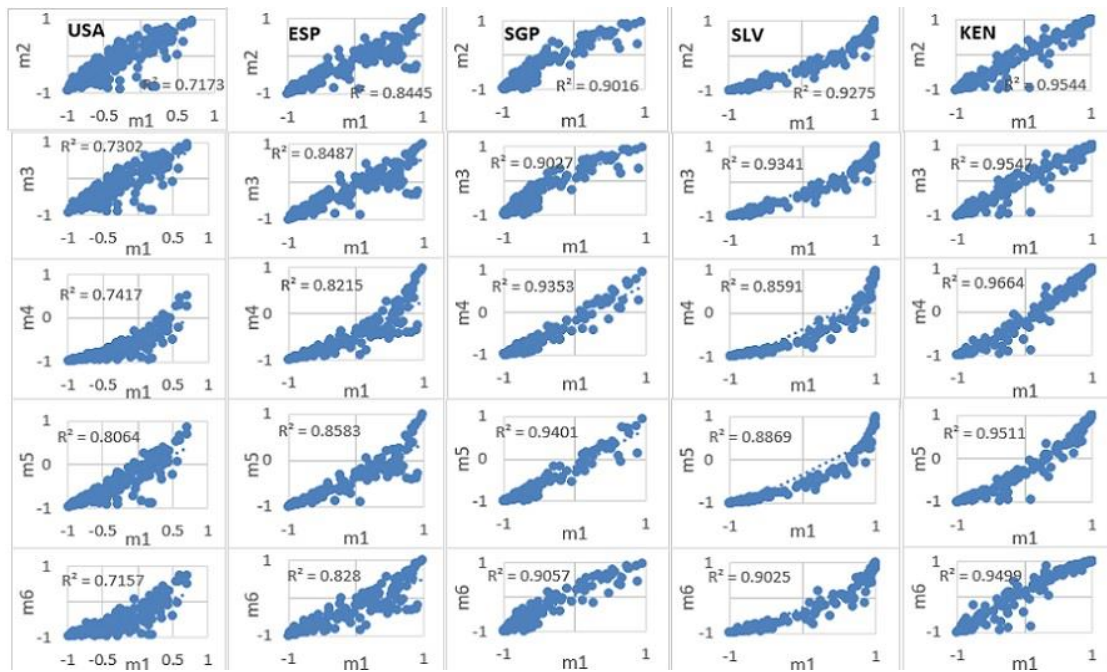

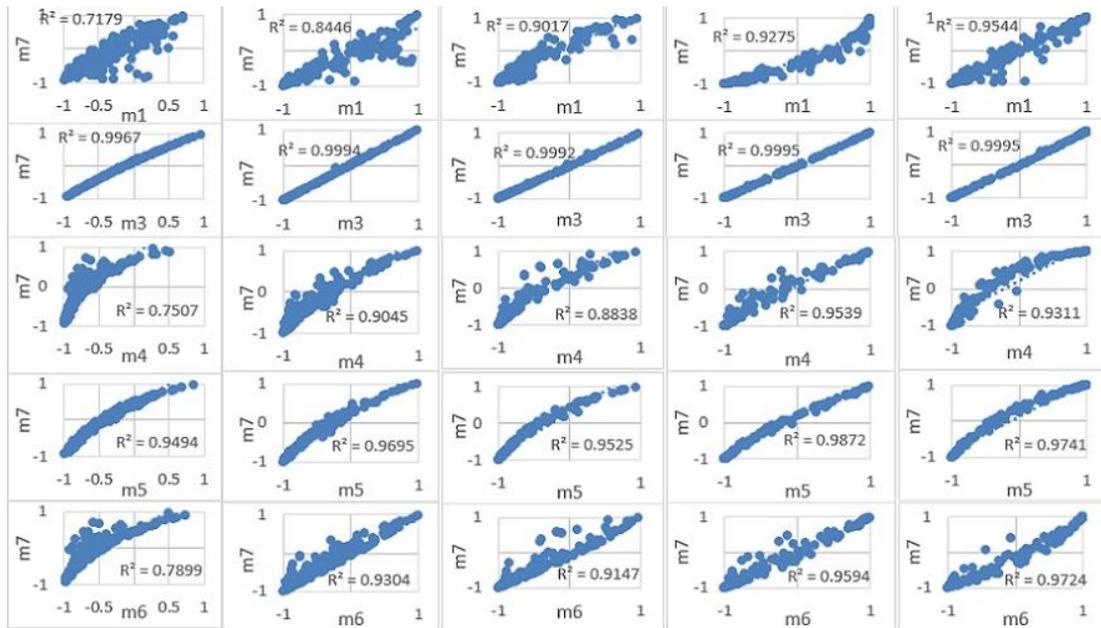

**Figure 3. Scatterplot of pairwise comparison of PAI variants.**

Comparison between methods shows that smaller countries/regions are especially more affected by non-overlapping method (M1) and that different types of production (all papers, domestic or international papers) in the diagonal might affect the results. But we do not find any particular pattern among methods to clearly differentiate what is the best one for highlighting preferred countries depending on the unit of analysis (co-authored papers vs co-authorship links). Instead, it would depend on the purposes of the study, in which one of them will be applied in order to mitigate the effect of a high proportion of domestic or international co-authorship links.

*4.2. Size of countries*

In order to explore the degree to which the indicators are size dependent, we carried out a similar comparative exercise to the one in the previous section (data and figures are available at the Supporting Information file). Table 5 shows Pearson and Spearman correlations for PAI in each method considering total number of papers or the number of international collaborative papers by countries. Note that when calculating correlation coefficients, we remove all countries whose PAI value is equal to -1 (as these indicate that there are no scientific collaborations between the two countries). As observed, most Pearson's correlations are quite low, regardless of large or small countries for both total publications' counts and international collaborative paper counts (below 0.2). There is hardly a linear correlation between PAI and the scientific size of countries. Spearman's correlations are higher, indicating that although they do not correlate, there is a relation between PAI and scientific size. Differences between Pearson and Spearman are observed for Spain, Slovenia, and Kenya. But there are no clear patterns in the set of countries under study and we are not able to generalize these results for all countries.

|  | Pearson | | | | | Spearman | | | | |
|---|---|---|---|---|---|---|---|---|---|---|
|  | **All papers** | | | | | | | | | |
|  | USA | ESP | SGP | SLV | KEN | USA | ESP | SGP | SLV | KEN |
| m1 | 0.08 | 0.02 | 0.23 | 0.01 | -0.14 | -0.03 | 0.29 | 0.05 | 0.32 | -0.59 |
| m2 | 0.23 | 0.12 | 0.20 | 0.00 | -0.15 | 0.03 | 0.36 | 0.04 | 0.31 | -0.59 |
| m3 | 0.21 | 0.09 | 0.04 | -0.09 | -0.23 | 0.02 | 0.35 | -0.10 | 0.21 | -0.63 |
| m4 | -0.07 | -0.03 | 0.18 | -0.03 | -0.18 | -0.17 | 0.22 | 0.04 | 0.31 | -0.59 |
| m5 | 0.17 | 0.06 | 0.05 | -0.08 | -0.21 | 0.01 | 0.34 | -0.13 | 0.19 | -0.63 |
| m6 | -0.07 | -0.03 | 0.23 | 0.01 | -0.14 | -0.22 | 0.18 | 0.05 | 0.32 | -0.59 |
| m7 | 0.22 | 0.12 | -0.16 | -0.04 | 0.19 | 0.02 | 0.36 | -0.04 | -0.27 | 0.61 |
|  | **Internationally collaborative papers** | | | | | | | | | |
|  | USA | ESP | SGP | SLV | KEN | USA | ESP | SGP | SLV | KEN |
| m1 | 0.05 | 0.06 | 0.20 | 0.03 | -0.17 | -0.02 | 0.28 | 0.05 | 0.28 | -0.56 |
| m2 | 0.20 | 0.17 | 0.18 | 0.02 | -0.18 | 0.04 | 0.35 | 0.04 | 0.27 | -0.56 |
| m3 | 0.17 | 0.14 | 0.02 | -0.07 | -0.26 | 0.03 | 0.34 | -0.08 | 0.17 | -0.60 |
| m4 | -0.07 | 0.01 | 0.15 | -0.02 | -0.22 | -0.14 | 0.22 | 0.04 | 0.26 | -0.56 |
| m5 | 0.14 | 0.10 | 0.03 | -0.05 | -0.23 | 0.02 | 0.34 | -0.12 | 0.16 | -0.60 |
| m6 | -0.07 | 0.02 | 0.20 | 0.03 | -0.17 | -0.19 | 0.19 | 0.05 | 0.28 | -0.56 |
| m7 | 0.19 | 0.17 | -0.15 | -0.07 | 0.22 | 0.04 | 0.35 | -0.04 | -0.23 | 0.59 |

**Table 5. Pearson (left) and Spearman (right) correlation between size of countries (total and in international collaboration) vs PAI results.**

## 5. DISCUSSION AND CONCLUSIONS

This study presents a review of the methods used in the scientometric literature for examining preferred partners in international collaboration. We provide a review of the related literature, set up a framework on how to use PAI in international scientific collaboration studies with a four-step procedure, and compare the results of its implementation within specific countries. In this way, we can identify which countries appear as preferred/strongest partners in each of the countries analyzed and what are the effects of applying different methods (overlapping, non-overlapping and self-exclusive strategies). We also present a quick exploration on the potential relationship between PAI and countries' scientific size, using their total number of papers and total number of internationally collaborative papers as measurements. However, we need to further investigate size-independent characteristic of the indicator in future research.

All previously defined PAI definition strategies are shown in Table 2, in which we can find that the non-overlapping method is only paired with zero main diagonal values, but for the overlapping methods discussed by Zitt et al. (2000), five strategies are provided, i.e., zero, an iterative way, total number of publications, total number of international collaborations, and total number of domestic collaborations. This does not mean that we cannot calculate a non-overlapping PAI with a non-zero main diagonal value; instead, future researchers can attempt to combine different definitions with distinct main diagonal values, and even normalization strategies. For instance, one will be able to define the non-overlapping PAI with an iterative strategy, but such action depends on the question under investigation.

For scientifically small countries, there should be small differences of applying overlapping and self-exclusive strategies, because the "exclusive" value is rather small compared with the other part in their formulas. However, for large countries, such differences cannot be ignored as should be considered carefully when one applies any of these strategies. This is the case for example with other indicators such as the activity index. For a more comprehensive review, we refer to (Rousseau, 2018). Therefore, we argue that researchers should determine the most appropriate indicator based upon their current situations.

In other hand, previous studies (e.g., Zitt et al., 2000) use two indices to measure network asymmetries in terms of the relative strength of scientific linkages in science —Affinity Index (AFI) and Probabilistic Affinity Index (PAI). AFI allows for the calibration of relative importance and asymmetries between countries, measuring the amount of collaborative papers published jointly and the total number of international collaborations of each country. These studies argued that AFI is size-dependent while PAI is not. However, such arguments do not clearly define what size-dependency is before being established. To be best of our knowledge, none of the previous literature has clearly defined mathematically what a size-dependent indicator is, and how to prove it step by step. In any case, this is not the focus of our paper which does not offer any additional contribution on the mathematic definition. Instead, we present a mini empirical study by quantifying the correlation between PAIs the scientific size of countries (measured by the total number of publications and total number of internationally collaborative publications of a country). We argue that PAIs have no linear relationship with scientific size, but do have some other types of relationships, per the differences between Pearson's and Spearman's correlations. Yet, again, further research is required on both theoretical and empirical sides to investigate size-dependency of PAI more accurately.

This study still requires further analysis in order to overcome other limitations and respond to other important questions related to the scientific capacities and socio-economic conditions of countries. At the methodological level, approaches with different counting methods (Park et al., 2016; Perianes-Rodriguez et al., 2016) will be analyzed to explore the effect of attributing coauthored publications as a full publication to each country or rather proportionally. Furthermore, the current paper purely applies PAI into country-country coauthorship networks. Future work can consider apply PAI into author- or institution-level coauthorship networks and even citation (and co-citation) networks. For instance, when utilizing PAI in an author co-citation network, one can know more details and have a deeper understanding on the dependency/independency of two certain authors in scientific intellectual structures. Finally, the analysis is conducted using bibliometric techniques and as always, the limitations and assumptions embedded in such analyses apply. Caution is therefore recommended in interpreting the findings.

## ACKNOWLEDGMENTS

Part of this paper has been presented in the 2018 Annual Meeting of the Society for Social Studies of Science (4S). The authors would like to thank Ludo Waltman for the fruitful discussions, the audiences in the 2018 4S annual meeting, and Cassidy Sugimoto's group members for their insightful suggestions/comments. Financial support from EUROPA INVESTIGACIÓN 2019 (EIN2019-102867) funded by the Ministerio de Ciencia, Innovación y Universidades (Spain) and I-LINK 2019 (LINKA20266) funded by the Consejo Superior de Investigaciones Científicas (Spain) is also acknowledged.

## SUPPORTING DATA

Supporting data can be found at: https://drive.google.com/file/d/1PoYVkBQ1aNRou4HFBCXCCmlvLM-v4afk/view?usp=sharing

## COMPETING FINANCIAL INTERESTS

The authors declare no competing financial interest.